# Impact maturity times and citation time windows: The 2-year maximum journal impact factor

P. Dorta-González [a*], M.I. Dorta-González [b]

[a] Departamento de Métodos Cuantitativos en Economía y Gestión, Universidad de Las Palmas de Gran Canaria, Gran Canaria, España; [b] Departamento de Estadística, Investigación Operativa y Computación, Universidad de La Laguna, Tenerife, España.

**ABSTRACT**

Journal metrics are employed for the assessment of scientific scholar journals from a general bibliometric perspective. In this context, the Thomson Reuters journal impact factors (JIF) are the citation-based indicators most used. The 2-year journal impact factor (*2-JIF*) counts citations to one and two year old articles, while the 5-year journal impact factor (*5-JIF*) counts citations from one to five year old articles. Nevertheless, these indicators are not comparable among fields of science for two reasons: *(i)* each field has a different impact maturity time, and *(ii)* because of systematic differences in publication and citation behaviour across disciplines. In fact, the *5-JIF* firstly appeared in the Journal Citation Reports (JCR) in 2007 with the purpose of making more comparable impacts in fields in which impact matures slowly. However, there is not an optimal fixed impact maturity time valid for all the fields. In some of them two years provides a good performance whereas in others three or more years are necessary. Therefore, there is a problem when comparing a journal from a field in which impact matures slowly with a journal from a field in which impact matures rapidly. In this work, we propose the 2-year maximum journal impact factor (*2M-JIF*), a new impact indicator that considers the 2-year rolling citation time window of maximum impact instead of the previous 2-year time window. Finally, an empirical application comparing *2-JIF*, *5-JIF*, and *2M-JIF* shows that the maximum rolling target window reduces the between-group variance with respect to the within-group variance in a random sample of about six hundred journals from eight different fields.

# 1. Introduction

This work is related to journal metrics and citation-based indicators for the assessment of scientific scholar journals from a general bibliometric perspective. During decades, the *journal impact factor* (JIF) has been an accepted indicator in ranking journals, however, there are increasing arguments against the fairness of using the JIF as the sole ranking criteria (Waltman & Van Eck, 2013).

The *2-year impact factor* published by Thomson Reuters in the *Journal Citation Reports* (JCR) is defined as the average number of references to each journal in a current year with respect to 'citable items' published in that journal during the two preceding years (Garfield, 1972). Since its formulation, the JIF has been criticized for some arbitrary decisions involved in its construction. The definition of 'citable items' (including letters and peer reviewed papers –articles, proceedings papers, and reviews–), the focus on the two preceding years as representation of impact at the research front, etc., have been discussed in the literature (Bensman, 2007; Moed et al., 2012) and have given rise to suggestions of many possible modifications and improvements (Althouse et al., 2009; Bornmann & Daniel, 2008). In response, Thomson Reuters has incorporated the *5-year impact factor*, the *eigenfactor score*, and the *article influence score* (Bergstrom, 2007) to the journals in the online version of the JCR since 2007. These journal indicators are most useful for comparing journals in the same subject category. In this respect, the 2-year and the 5-year impact factor lead statistically to the same ranking (Leydesdorff, 2009; Rousseau, 2009). Yet, it seems that in many cases, but not always, the 5-year impact factor is larger than the 2-year one (Rousseau, 2009).

However, these indicators do not solve the problem when comparing journals from different fields of science. Different scientific fields have different citation practices. Citation-based bibliometric indicators need to be normalized for such differences between fields, in order to allow for meaningful between-field comparisons of citation impact. This problem of field-specific differences in citation impact indicators comes from institutional research evaluation (Leydesdorff & Bornmann, 2011; Van Raan et al., 2010). Institutes are populated by scholars with different disciplinary backgrounds and research institutes often have among their missions the objective of integrating interdisciplinary bodies of knowledge (Leydesdorff & Rafols, 2011; Wagner et al., 2011).

There are statistical patterns which are field-specific and allow for the normalization of the JIF. Garfield (1979) proposes the term 'citation potential' for systematic differences among fields of science based on the average number of references. For example, in the biomedical fields long reference lists with more than fifty items are common, but in mathematics short lists with fewer than twenty references are the standard (Dorta-González & Dorta-González, 2013). These differences are a consequence of the citation cultures, and can lead to significant differences in the JIF across fields of science because the probability of being cited is affected. In this sense, this is the factor that has most frequently been used in the literature to justify the differences between fields of science, as well as the most employed in source-normalization (Leydesdorff & Bornmann, 2011; Moed, 2010; Zitt & Small, 2008).

However, the variables that to a greater degree explain the variance in the impact factor do not include the average number of references (Dorta-González & Dorta-González, 2013) and therefore it is necessary to consider some other sources of variance in the normalization process such as the ratio of references to journals included in the JCR, the field growth, the ratio of JCR references to the target window and the proportion of cited to citing items. Given these large differences in citation practices, the development of bibliometric indicators that allow for meaningful between-field comparisons is clearly a critical issue (Waltman & Van Eck, 2013).

Traditionally, normalization for field differences has usually been done based on a field classification system. In this approach, each publication belongs to one or more fields and the citation impact of a publication is calculated relative to the other publications in the same field. Most efforts to classify journals in terms of fields of science have focused on correlations between citation patterns in core groups assumed to represent scientific specialties (Leydesdorff, 2006; Rosvall & Bergstrom, 2008). Indexes such as the *JCR subject category list* accommodate a multitude of perspectives by listing journals under different groups (Pudovkin & Garfield, 2002; Rafols & Leydesdorff, 2009). In this sense, Egghe & Rousseau (2002) propose the *aggregate impact factor* in a similar way as the JIF, taking all journals in a category as a meta-journal. However, the position of individual journals of merging specialties remains difficult to determine with precision and some journals are assigned to more than one category. Moreover, the delineation between fields of science and next-lower level specialties has until now remained an unsolved problem in bibliometrics because these delineations are fuzzy at each moment of time and develop dynamically over time. Therefore, classifying a dynamic system in terms of fixed categories can be expected to lead to error because the classification system is then defined historically while the dynamics of science is evolutionary (Leydesdorff, 2012, p.359).

Recently, the idea of source normalization was introduced, which offers an alternative approach in normalizing field differences. In this approach, normalization is done by looking at the referencing behaviour of citing journals. Journal performance is a complex multi-dimensional concept difficult to be fully captured in one single metric (Moed et al., 2012, p. 368). This resulted in the creation of many other quality metric indices such as the *fractionally counted impact factor* (Leydesdorff & Bornmann, 2011), *audience factor* (Zitt & Small, 2008), *source normalized impact per paper* (Moed, 2010), *scimago journal ranking* (González-Pereira et al., 2009) and *central area index* (Dorta-González & Dorta-González, 2010, 2011) to name a few. All these metrics have their merits, but none include any great degree of normalization in relation to the citation maturity time.

Impact indicators have varying publication and citation periods and the chosen length of these periods enables a distinction between synchronous and diachronous impact indicators. To collect data for calculations of diachronous journal impact factors several volumes of the JCR are needed (Frandsen & Rousseau, 2005). The term diachronous refers to the fact that the data used to calculate is derive from a number of different years with a starting point somewhere in the past and encompassing subsequent years. However, these indicators are not going into the subject of relative impact or normalizations (Frandsen & Rousseau, 2005).

Although journal impact factors can be considered historically as the first way of trying to normalize citation distributions by using averages over 2 years (Leydesdorff, 2009), it has been recognized that citation distributions vary among fields of science and that this needs to be normalized. This is the motivation in considering the two years of maximum citations and variable time windows in providing an alternative to the current journal impact factor.

In this paper, we provide a source normalization approach based on variable citation time windows and we empirically compare this with the traditional normalization approach based on a fixed target window. We propose the *2-year maximum journal impact factor* (*2M-JIF*), a new impact indicator that considers the 2-year target window of maximum impact instead of the previous 2-year target window. This new indicator is intuitively simple, allows for statistical testing, and accords with the state of the art.

In order to compare this new impact indicator with the 2-year and 5-year time window impact factors, an empirical application with about six hundred journals belonging to eight different subject categories is presented. As the main conclusion, we obtain that our indicator reduces the between-group variance in relation to the within-group variance.

The organization of this paper is as follows. ''The fixed citation time window'' discusses the issue of the selection of the journals optimal citation time window. ''The variable citation time window'' introduces the new bibliometric indicator that we are studying. ''Results and discussion'' presents the results of our empirical analysis. Finally, ''Conclusions'' summarizes our conclusions.

**2. The fixed citation time window**

A journal impact indicator is a measure of the number of times papers published in a census period cite papers published during an earlier target window.

*2.1 The 2-year and 5-year citation time windows*

The *2-JIF* reported by Thomson Reuters considers a one year census period and uses the previous two years as the target window. As an average, the *2-JIF* is based on two elements: the numerator, which is the number of citations in the current year to any items published in a journal in the previous two years, and the denominator, which is the number of 'citable items' (articles, proceedings papers, reviews, and letters) published in those same two years (Garfield, 1972). Journal items include 'citable items', but also editorials, news, corrections, retractions, and other items. Similarly, the *5-JIF* considers a one year census period and uses the previous five years as the target window.

Let $NArt_t^i$ be the number of 'citable items' in year *t* of journal *i*. Let $NCit_{t,t-j}^i$ be the number of times in year *t* that the year *t-j* volumes of journal *i* are cited by journals in the JCR. Then, the n-year impact factor in year *t* of journal *i* is defined as:

$$n\text{-}JIF_t^i = \sum_{j=1}^{n} NCit_{t,t-j}^i \Big/ \sum_{j=1}^{n} NArt_{t-j}^i,$$

and in the specific cases of two and five years,

$$2\text{-}JIF_t^i = \frac{NCit_{t,t-1}^i + NCit_{t,t-2}^i}{NArt_{t-1}^i + NArt_{t-2}^i}$$

and

$$5\text{-}JIF_t^i = \sum_{j=1}^{5} NCit_{t,t-j}^i \Big/ \sum_{j=1}^{5} NArt_{t-j}^i.$$

Nevertheless, a common source of variance in the *n*-year impact factor is the citation potential, a measure of the citation characteristics of the field where the journal is positioned, determined by how often and how rapidly authors cite other works, and how well their field is covered by the database. The 'citation potential' can be conceived as a measure of the field's topicality (Moed et al., 2012). Fields with a high topicality tend to attract many authors who share an intellectual interest, and in this sense can be qualified as 'popular'. Developments in these fields move quickly. Papers are written in a limited number of highly visible journals, and authors tend to cite, apart from the common intellectual base, the most recent papers of their colleagues. These popular fields will tend to have a higher *2-JIF* (Moed et al., 2012). In this paper we will refer to journals in these popular fields as journals with rapid *impact maturity time*.

Therefore, there is not an optimal fixed *n*-value valid for all the journals and fields. In some cases two years provide a good measure of performance but in others three or more years is necessary.

*2.2 The 3-year citation time window*

The impact indicator reported by Elsevier SciVerse Scopus considers a one year census period and uses the previous three years as the target window. The numerator is the number of citations in the current year to any items published in a journal in the previous three years, and the denominator is the number of 'refereed items' (articles, proceedings papers, and reviews) published in the same three years.

However, this intermediate citation time window is not a solution because in the intention to be valid for all journals and fields it is really not a good measure both for fields with slow and rapid citation maturity times.

*2.3 The complete citation time window*

In addition to the static variance (in each yearly JCR) the dynamic variance can be reduced by using total citations (i.e., the complete citation window) instead of the window of the last *n* years. However, this model does not improve on the model using the previous 2-year time window (Leydesdorff & Bornmann, 2011, p.228). Therefore, a focus on the last two years (following Garfield's (1972) suggestion to measure the 'research front') works better than including the complete historical record, i.e., 'total cites'.

## 3. The variable citation time window

As we have introduced in the previous section, a journal impact indicator is a measure of the number of times that papers published in a census period cite papers published during an earlier target window. However, relevant citation windows can vary both among fields and over time. Therefore, although fixed citation time windows have been considered in the literature for decades, there is no evidence in justifying its suitability in relation to a variable time window.

The delimitation among fields of science and the next-lower level specialties has until now remained an unsolved problem in bibliometrics because these delineations are fuzzy at each moment in time and develop dynamically over time. For this reason, it is not recommended selecting the target window in relation to the subject category in which the journal is included.

Researchers in the fields where impact matures rapidly have an immediate 'consumption' (diffusion and use) of the scientific production (e.g., Biomedicine and Computer Science). Conversely, in fields where impact matures slowly researchers have a slow 'consumption' of the scientific production (e.g., Mathematics and Economics).

There is not an optimal fixed impact maturity time valid for all journals. The choice for a variable rather than a fixed (2, 3, or 5 years) citation time window is based on the observation that in many fields citations have not yet peaked after 2 years, and in other fields citations have peaked long before 5 years. Therefore, the application of a 2-year variable window is the optimal compromise for fields in which impact matures slowly in reaching its maximum citations while not penalising fields in which impact matures rapidly.

Figure 1 shows the citations distribution of four journals with different performance. Journals *A* and *C* belong to a field in which impact matures rapidly, while journals *B* and *D* belong to a field in which impact matures slowly. The citations are the numerators in the impact formula, therefore if these journals have the same size, that is they have published the same number of papers in the last years, then *A* has greater impact than *C*, and *B* has greater impact than *D*. Nevertheless, which journal has greater impact, *A* or *B*? And, which journal has greater impact, *C* or *D*?

[Figure 1 about here]

In this work, we propose the *2-year maximum journal impact factor* (*2M-JIF*), a new impact indicator that considers the 2-year citation time window of maximum impact instead of the previous two years. The idea is to consider, for each journal, the citation time window with the highest average number of citations (i.e., the most advantageous period for each journal).

We define the *rolling impact factors* in year $t$ of journal $i$ as:

$$R_j\text{-}JIF_t^i = \frac{NCit_{t,\,t-j}^i + NCit_{t,\,t-j-1}^i}{NArt_{t-j}^i + NArt_{t-j-1}^i},\; j=1,...,h,$$

and the *2-year maximum journal impact factor* in year $t$ of journal $i$ as following:

$$2M\text{-}JIF_t^i = max_{j=1,...,h}\left\{R_j\text{-}JIF_t^i\right\} = max_{j=1,...,h}\left\{\frac{NCit_{t,\,t-j}^i + NCit_{t,\,t-j-1}^i}{NArt_{t-j}^i + NArt_{t-j-1}^i}\right\}.$$

We define the *impact maturity time* in year $t$ of a journal $i$ as the number of years from $t$ to that in which the maximum impact is achieved, that is, if $2M\text{-}JIF_t^i = R_j\text{-}JIF_t^i$ then $j+1$ is the impact maturity time in year $t$ of journal $i$.

Note that $R_1\text{-}JIF_t^i = 2\text{-}JIF_t^i$ and, therefore, if $2M\text{-}JIF_t^i = R_1\text{-}JIF_t^i$ then two is the impact maturity time in year $t$ of journal $i$.

In the particular case of a journal publishing the same number of articles per year, that is $NArt^i = NArt_{t-1}^i = \cdots = NArt_{t-h-1}^i$, then

$$2M\text{-}JIF_t^i = \frac{1}{2NArt^i} max_{j=1,...,h}\left\{NCit_{t,\,t-j}^i + NCit_{t,\,t-j-1}^i\right\}.$$

Consider the example in Figure 2 with a journal $A$ from a field in which impact matures rapidly and a journal $B$ from a field in which impact matures slowly. If these journals have the same size $NArt = NArt^A = NArt^B$, then:

$$2M\text{-}JIF_t^A = R_1\text{-}JIF_t^A = \frac{NCit_{t,\,t-1}^A + NCit_{t,\,t-2}^A}{2NArt} = 2\text{-}JIF_t^A,$$

$$2M\text{-}JIF_t^B = R_5\text{-}JIF_t^B = \frac{NCit_{t,\,t-5}^B + NCit_{t,\,t-6}^B}{2NArt} > 2\text{-}JIF_t^B.$$

In this example, considering a 2-year fixed citation time window penalizes journal $B$ in its assessment with $A$, and it seems better to consider the 2-year maximum citation time window.

[Figure 2 about here]

In order to compare *2-JIF* and *2M-JIF*, consider the citation distribution of journal $B$ in Figure 3. Note that:

$$2M\text{-}JIF = \frac{a+c+b+d}{2NArt} = \frac{a+b}{2NArt} + \frac{c+d}{2NArt} = 2\text{-}JIF + \frac{c+d}{2NArt}.$$

Then, $\frac{c+d}{2NArt}$ is the observed but not measured impact by the *2-JIF*.

[Figure 3 about here]

Finally, the merit and quality of an indicator depends on its statistical properties, validity, and reliability. In this context, the validity is related to the capacity of measuring the impact of the journal at the research front, and the reliability is related to the ability of measuring the impact with minimum errors. In this sense, the statistical properties of this new indicator are checked in the following empirical application.

## 4. Methods and Materials

The underlying bibliometric data in the empirical application was obtained from the online version of the 2011 *Journal Citation Reports* (JCR) Science edition during the first week of november 2012. The JCR database (reported by Thomson Reuters – ISI, Philadelphia, USA) is available at the website www.webofknowledge.com. In the JCR, journals are assigned by Thomson Reuters' experts into one or more journal categories according to cited and citing relationships with the journals in the categories. The journal categories, also referred as *subject category list,* are treated as fields and subfields of science.

In the comparative analysis among *2-JIF*, *5-JIF*, and *2M-JIF*, one journal category from each of the eight clusters obtained by Dorta-González & Dorta-González (2013) were considered. This was done in order to obtain journals with systematic differences in publication and citation behaviour. A total of 618 journals were considered in this empirical application. The categories and the number of journals are as follows: Astronomy & Astrophysics (56); Biology (85); Ecology (134); Engineering, Aerospace (27); History & Philosophy of Science (56); Mathematics, Interdisciplinary Applications (92); Medicine, Research & Experimental (112); and Multidisciplinary Sciences (56).

## 5. Results and discussion

In the empirical application we study which citation time window in the impact indicator produces a closer data distribution among scientific fields in relation to its centrality and variability measures.

*5.1 A sample of 24 journals in eight JCR categories*

Table 1 shows a sample of 24 randomly selected journals from those with the greatest overall impact (total citations) in eight JCR categories from the different clusters identified by Dorta-González & Dorta-González (2013). This was done in order to obtain journals with systematic differences in publication and citation behaviour. This table contains the citations in year 2011 of items in the period 2006-2010 and the number of 'citable items' in the same period. Notice the important differences in publications and citations among journals and fields. This variance in the data is very relevant in the impact factors. In particular, in the number of publications, note an exponential increase in *PLOS ONE*, and a linear decrease in *ANN NY ACAD SCI* and *LIFE SCI*.

[Table 1 about here]

For the journals considered, Table 2 shows some journal impact factors with different citation time windows. Citation maturity time varies from one category to another (between two and five years). Notice the ampleness in the interval of variation for each indicator. The $R_1$ varies from 0.667 to 15.748, while *2M-JIF* varies from 0.818 to 18.335, for example. The general pattern is an increment in the *2M-JIF*. However, this increment is in percentage terms much higher in the smaller values. For example, note the case *J GUID CONTROL DYNAM* where *2M-JIF* is about 45% higher than $R_1$. By contrast, notice the cases *P NATL ACAD SCI USA* and *TRENDS ECOL EVOL* in which *2M-JIF* is just 15% higher than $R_1$. This effect produces a concentration of data in the case of *2M-JIF,* and consequently a reduction in the variance.

[Table 2 about here]

*5.2 A sample of 618 journals in eight JCR categories*

Table 3 provides the Pearson rank correlations for all pairs of indicators, both for journal categories and aggregate data. The general pattern that can be observed based on the correlations reported in this table, is that the five impact indicators are all quite strongly correlated, with most of the Pearson correlations above 0.90. The correlations of the four rolling indicators with the maximum indicator are somewhat higher, both within each category and the aggregate data, but the difference is not large. Moreover, the *2M-JIF* can explain more than 90% of the variance in the rolling indicators, $r^2=0.95^2=0.90$. However, one should be careful when drawing conclusions from the correlations reported in this table. The different indicators all have skewed distributions, with many journals with relatively low indicator values and only a small number of journals with high indicator values. These skewed distributions fairly easily give rise to high Pearson correlations.

[Table 3 about here]

The number of journals in which the rolling impact factor is the maximum value is showed in Table 4. Note there is not a valid optimal impact maturity time for all fields. In some cases two years provide a good measure of performance but in others three or more years is necessary. Note that impact matures rapidly (two years) in *Astronomy & Astrophysics*, followed by *Medicine, Research & Experimental* (three years). Impact matures slowly in *Ecology* and *Mathematics, Interdisciplinary Applications* (five years). The remaining fields are in intermediate situations, from four to five years.

[Table 4 about here]

Central-tendency and variability measures for the eight JCR categories analysed are showed in Table 5. All the indicators have skewed distributions, with many journals with relatively low indicator values and only a small number of journals with high indicator values. This is the reason why in these skewed distributions medians are well below means in all cases. Note high differences between categories in medians, means, and standard deviations.

The general pattern is an increment in the *2M-JIF* with respect to the rolling indicators. However, this increment is in percentage terms much higher in the smaller values. For example, in *History & Philosophy of Science* the median in *2M-JIF* is around a 60% higher than $R_1$ (50% in the case of the mean). By contrast, in *Medicine, Research & Experimental* the median in *2M-JIF* is around 15% higher than $R_1$ (25% in the case of the mean). This effect produces a concentration of data and consequently a reduction in the variance when considering the *2M-JIF*. As a specific case, note the mean is over four times the median in *Multidisciplinary Sciences*. This is also observed in the very large standard deviation.

[Table 5 about here]

Finally, we will test if the maximum citation time window reduces the between-group variance in relation to the within-group variance. Table 6 shows the central-tendency measures for the aggregate data. It also shows the between-group variances. Note that all target windows reduce the between-group variance. However, the maximum citation time window produces the greatest reduction (3.203). Thus, this normalization by means of variable target windows reduces the between-group variance over 80%, when compared to within-group variance.

[Table 6 about here]

## 6. Conclusions

Different scientific fields have different citation practices. Citation-based bibliometric indicators need to be normalized for such differences between fields in order to allow for meaningful between-field comparisons of citation impact. In this paper, we provide a source normalization approach, based on a variable target window and we compare it with a traditional normalization approach based on a fixed target window.

An empirical application, with about six hundred journals from eight different fields, shows that our maximum citation time window reduces the between-group variance in relation to the within-group variance more than the rest of the indicators analyzed.

The journal categories considered are in very different areas in relation to the impact maturity time. Some of them are penalized by the *2-JIF* and favored by the *5-JIF*, and vice versa. This is the main reason why it is necessary to be cautious when comparing journal impact factors from different fields. In this sense, our index has behaved well in a great number of journals from very different fields.

Finally, we have not empirical evidences about the optimality in using a time window of two years instead of some other value. We think that perhaps this aspect could be an interesting issue for a future work.

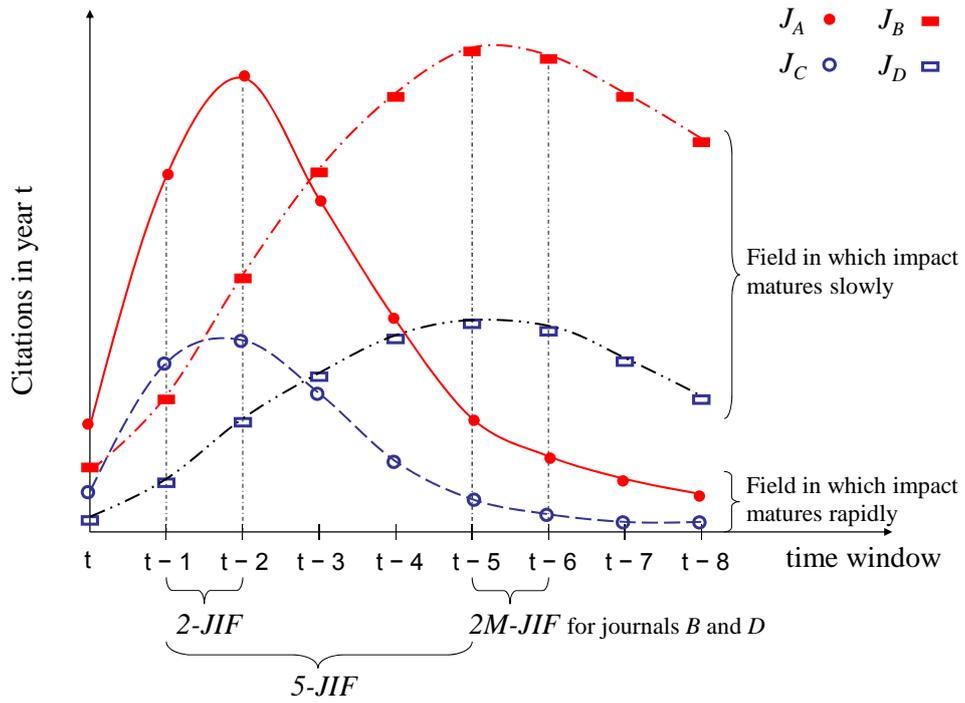

Figure 1: Citations distribution and impact measures of journals from fields in which impact matures rapidly (*A* with greater impact than *C*) and slowly (*B* with greater impact than *D*)

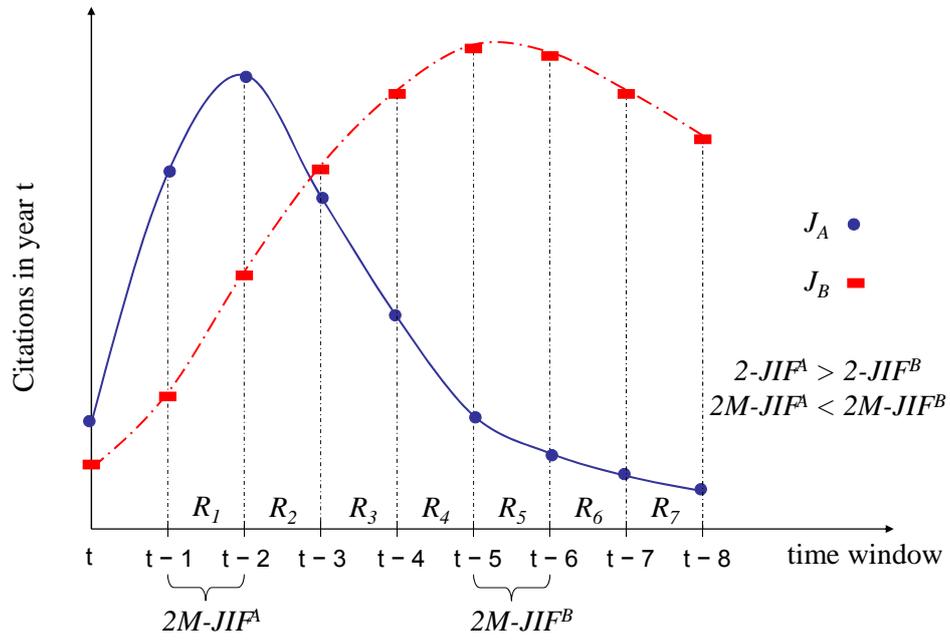

Figure 2: Two journals with different impact maturity time.

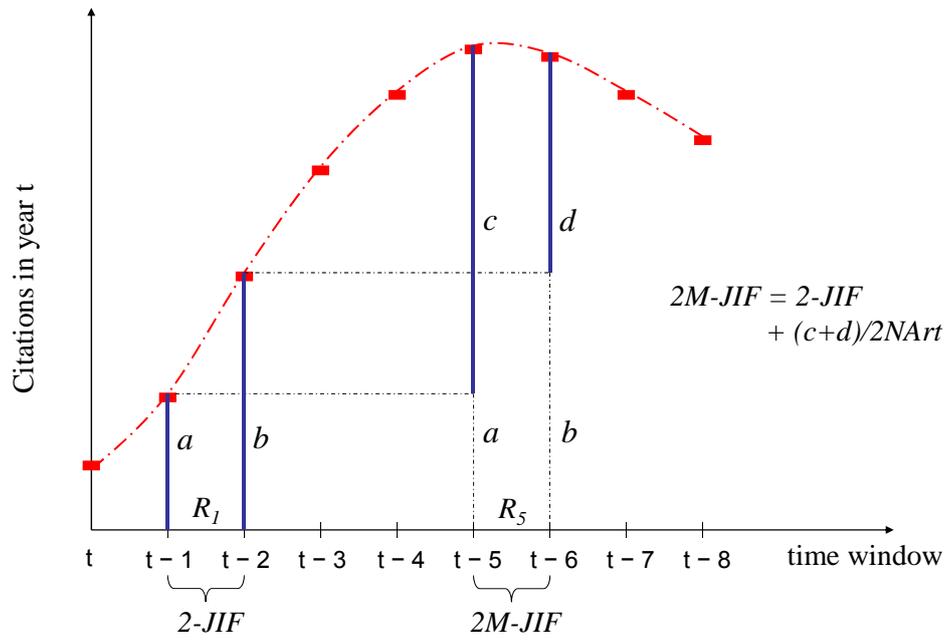

Figure 3: Comparing *2-JIF* and *2M-JIF* in a journal publishing *NArt* papers per year.

Table 1: A sample of 24 randomly selected journals from those with greatest overall impact (total citations) in eight very different JCR categories

| Abbreviated journal title | Category | $NCit^i_{2011,t}$ | | | | | $NArt^i_t$ | | | | |
|---|---|---|---|---|---|---|---|---|---|---|---|
| | | 2010 | 2009 | 2008 | 2007 | 2006 | 2010 | 2009 | 2008 | 2007 | 2006 |
| AIAA J | EA | 239 | 354 | 474 | 418 | 467 | 275 | 286 | 301 | 311 | 356 |
| AM NAT | E | 663 | 1052 | 1028 | 1159 | 1003 | 171 | 192 | 190 | 197 | 179 |
| ANN NY ACAD SCI | MS | 2505 | 3382 | 3827 | 2947 | 3193 | 702 | 1164 | 975 | 1034 | 1415 |
| ASTRON ASTROPHYS | A&A | 8657 | 8330 | 6992 | 7174 | 6270 | 1916 | 1787 | 1789 | 1977 | 1935 |
| ASTROPHYS J | A&A | 14641 | 17267 | 12160 | 11738 | 10412 | 2501 | 2796 | 2128 | 2848 | 2707 |
| BIOL PHILOS | H&PS | 66 | 29 | 39 | 59 | 49 | 39 | 40 | 36 | 35 | 28 |
| BIOMETRIKA | B | 103 | 203 | 222 | 246 | 225 | 79 | 81 | 75 | 74 | 79 |
| BRIT J PHILOS SCI | H&PS | 27 | 41 | 59 | 38 | 45 | 31 | 31 | 32 | 32 | 28 |
| ECOLOGY | E | 1292 | 2073 | 2317 | 2227 | 2237 | 357 | 337 | 345 | 317 | 333 |
| ECONOMETRICA | MIA | 136 | 239 | 228 | 326 | 373 | 65 | 61 | 47 | 51 | 53 |
| EXP HEMATOL | MR&E | 308 | 485 | 627 | 644 | 570 | 127 | 146 | 172 | 214 | 194 |
| FASEB J | B | 2348 | 2633 | 2845 | 2655 | 3200 | 462 | 410 | 412 | 388 | 486 |
| HIST SCI | H&PS | 9 | 15 | 12 | 12 | 10 | 17 | 19 | 14 | 17 | 16 |
| IEEE T AERO ELEC SYS | EA | 124 | 163 | 216 | 270 | 302 | 136 | 126 | 128 | 133 | 117 |
| J ECONOMETRICS | MIA | 156 | 165 | 435 | 541 | 448 | 139 | 99 | 161 | 176 | 124 |
| J GUID CONTROL DYNAM | EA | 151 | 213 | 261 | 268 | 208 | 187 | 200 | 183 | 203 | 177 |
| LIFE SCI | MR&E | 538 | 675 | 883 | 1364 | 1919 | 228 | 252 | 289 | 498 | 702 |
| P NATL ACAD SCI USA | MS | 31558 | 41331 | 39642 | 38547 | 35707 | 3764 | 3765 | 3508 | 3494 | 3306 |
| P ROY SOC A-MATH PHY | MS | 397 | 346 | 323 | 453 | 359 | 183 | 194 | 175 | 197 | 196 |
| PHYS REV D | A&A | 13330 | 12498 | 11508 | 8183 | 7528 | 2854 | 2813 | 2863 | 2268 | 2375 |
| PLOS ONE | B | 22741 | 22780 | 15676 | 7041 | 765 | 6722 | 4403 | 2717 | 1230 | 137 |
| STRUCT EQU MODELING | MIA | 99 | 193 | 98 | 308 | 374 | 31 | 31 | 30 | 29 | 28 |
| TRENDS ECOL EVOL | E | 965 | 1476 | 1527 | 1468 | 1594 | 75 | 80 | 92 | 89 | 78 |
| VACCINE | MR&E | 3729 | 4702 | 3787 | 3536 | 3182 | 1105 | 1134 | 905 | 1046 | 928 |

*JCR categories: A&A Astronomy & Astrophysics; B Biology; E Ecology; EA Engineering, Aerospace; H&PS History & Philosophy of Science; MIA Mathematics, Interdisciplinary Applications; MR&E Medicine, Research & Experimental; MS Multidisciplinary Sciences.*

Table 2: Journal impact factors with different citation time windows for journals with different impact maturity times

| Abbreviated journal title | Category | $R_1=$ 2-JIF | $R_2$ | $R_3$ | $R_4$ | 2M-JIF | 5-JIF | Impact maturity time |
|---|---|---|---|---|---|---|---|---|
| AIAA J | EA | 1.057 | 1.411 | 1.458 | 1.327 | 1.458 | 1.277 | 4 |
| AM NAT | E | 4.725 | 5.445 | 5.651 | 5.750 | 5.750 | 5.280 | 5 |
| ANN NY ACAD SCI | MS | 3.155 | 3.370 | 3.372 | 2.507 | 3.372 | 2.997 | 4 |
| ASTRON ASTROPHYS | A&A | 4.587 | 4.285 | 3.762 | 3.437 | 4.587 | 3.979 | 2 |
| ASTROPHYS J | A&A | 6.024 | 5.976 | 4.803 | 3.987 | 6.024 | 5.102 | 2 |
| BIOL PHILOS | H&PS | 1.203 | 0.895 | 1.380 | 1.714 | 1.714 | 1.360 | 5 |
| BIOMETRIKA | B | 1.913 | 2.724 | 3.141 | 3.078 | 3.141 | 2.575 | 4 |
| BRIT J PHILOS SCI | H&PS | 1.097 | 1.587 | 1.516 | 1.383 | 1.587 | 1.364 | 3 |
| ECOLOGY | E | 4.849 | 6.437 | 6.864 | 6.868 | 6.868 | 6.007 | 5 |
| ECONOMETRICA | MIA | 2.976 | 4.324 | 5.653 | 6.721 | 6.721 | 4.700 | 5 |
| EXP HEMATOL | MR&E | 2.905 | 3.497 | 3.293 | 2.975 | 3.497 | 3.088 | 3 |
| FASEB J | B | 5.712 | 6.664 | 6.875 | 6.699 | 6.875 | 6.340 | 4 |
| HIST SCI | H&PS | 0.667 | 0.818 | 0.774 | 0.667 | 0.818 | 0.699 | 3 |
| IEEE T AERO ELEC SYS | EA | 1.095 | 1.492 | 1.862 | 2.288 | 2.288 | 1.680 | 5 |
| J ECONOMETRICS | MIA | 1.349 | 2.308 | 2.896 | 3.297 | 3.297 | 2.496 | 5 |
| J GUID CONTROL DYNAM | EA | 0.941 | 1.238 | 1.370 | 1.253 | 1.370 | 1.159 | 4 |
| LIFE SCI | MR&E | 2.527 | 2.880 | 2.855 | 2.736 | 2.880 | 2.732 | 3 |
| P NATL ACAD SCI USA | MS | 9.681 | 11.133 | 11.167 | 10.920 | 11.167 | 10.472 | 4 |
| P ROY SOC A-MATH PHY | MS | 1.971 | 1.813 | 2.086 | 2.066 | 2.086 | 1.987 | 4 |
| PHYS REV D | A&A | 4.558 | 4.229 | 3.838 | 3.384 | 4.558 | 4.027 | 2 |
| PLOS ONE | B | 4.092 | 5.401 | 5.756 | 5.710 | 5.756 | 4.537 | 4 |
| STRUCT EQU MODELING | MIA | 4.710 | 4.770 | 6.881 | 11.965 | 11.965 | 7.195 | 5 |
| TRENDS ECOL EVOL | E | 15.748 | 17.459 | 16.547 | 18.335 | 18.335 | 16.981 | 5 |
| VACCINE | MR&E | 3.766 | 4.163 | 3.753 | 3.403 | 4.163 | 3.700 | 3 |

*JCR categories: A&A Astronomy & Astrophysics; B Biology; E Ecology; EA Engineering, Aerospace; H&PS History & Philosophy of Science; MIA Mathematics, Interdisciplinary Applications; MR&E Medicine, Research & Experimental; MS Multidisciplinary Sciences.*

Table 3: Pearson rank correlations for all pairs of indicators

| Category | # Journals | | $R_2$ | $R_3$ | $R_4$ | 2M-JIF |
|---|---|---|---|---|---|---|
| Astronomy & Astrophysics | 56 | $R_1$ | 0.96 | 0.93 | 0.92 | 0.95 |
| | | $R_2$ | | 0.94 | 0.91 | 0.96 |
| | | $R_3$ | | | 0.88 | 0.98 |
| | | $R_4$ | | | | 0.89 |
| Biology | 85 | $R_1$ | 0.977 | 0.93 | 0.94 | 0.97 |
| | | $R_2$ | | 0.98 | 0.96 | 0.99 |
| | | $R_3$ | | | 0.98 | 0.98 |
| | | $R_4$ | | | | 0.97 |
| Ecology | 134 | $R_1$ | 0.99 | 0.98 | 0.97 | 0.99 |
| | | $R_2$ | | 0.98 | 0.95 | 0.98 |
| | | $R_3$ | | | 0.97 | 0.98 |
| | | $R_4$ | | | | 0.97 |
| Engineering, Aerospace | 27 | $R_1$ | 0.95 | 0.83 | 0.83 | 0.92 |
| | | $R_2$ | | 0.91 | 0.90 | 0.95 |
| | | $R_3$ | | | 0.98 | 0.95 |
| | | $R_4$ | | | | 0.95 |
| History & Philosophy of Science | 56 | $R_1$ | 0.89 | 0.82 | 0.85 | 0.89 |
| | | $R_2$ | | 0.93 | 0.83 | 0.90 |
| | | $R_3$ | | | 0.92 | 0.95 |
| | | $R_4$ | | | | 0.97 |
| Mathematics, Interdisciplinary Applications | 92 | $R_1$ | 0.91 | 0.81 | 0.77 | 0.86 |
| | | $R_2$ | | 0.92 | 0.82 | 0.90 |
| | | $R_3$ | | | 0.90 | 0.94 |
| | | $R_4$ | | | | 0.93 |
| Medicine, Research & Experimental | 112 | $R_1$ | 0.90 | 0.80 | 0.76 | 0.93 |
| | | $R_2$ | | 0.94 | 0.89 | 0.92 |
| | | $R_3$ | | | 0.96 | 0.90 |
| | | $R_4$ | | | | 0.87 |
| Multidisciplinary Sciences | 56 | $R_1$ | 0.96 | 0.91 | 0.91 | 0.96 |
| | | $R_2$ | | 0.97 | 0.94 | 0.98 |
| | | $R_3$ | | | 0.94 | 0.98 |
| | | $R_4$ | | | | 0.95 |
| Total | 618 | $R_1$ | 0.97 | 0.93 | 0.91 | 0.96 |
| | | $R_2$ | | 0.97 | 0.94 | 0.97 |
| | | $R_3$ | | | 0.96 | 0.97 |
| | | $R_4$ | | | | 0.95 |

$R_1$=2-JIF

Table 4: Number of journals in which the rolling impact factor is the maximum value

| Category | # Journals | $R_1$=2-JIF | $R_2$ | $R_3$ | $R_4$ |
|---|---|---|---|---|---|
| Astronomy & Astrophysics | 56 | 22 | 17 | 11 | 6 |
| | | 39.3% | 30.4% | 19.6% | 10.7% |
| Biology | 85 | 13 | 25 | 28 | 19 |
| | | 15.3% | 29.4% | 32.9% | 22.4% |
| Ecology | 134 | 7 | 31 | 41 | 55 |
| | | 5.2% | 23.1% | 30.6% | 41.0% |
| Engineering, Aerospace | 27 | 4 | 7 | 8 | 8 |
| | | 14.8% | 25.9% | 29.6% | 29.6% |
| History & Philosophy of Science | 56 | 12 | 16 | 12 | 16 |
| | | 21.4% | 28.6% | 21.4% | 28.6% |
| Mathematics, Interdisciplinary Applications | 92 | 10 | 22 | 22 | 38 |
| | | 10.9% | 23.9% | 23.9% | 41.3% |
| Medicine, Research & Experimental | 112 | 22 | 46 | 22 | 22 |
| | | 19.6% | 41.1% | 19.6% | 19.6% |
| Multidisciplinary Sciences | 56 | 13 | 14 | 18 | 11 |
| | | 23.2% | 25.0% | 32.1% | 19.6% |
| Total | 618 | 103 | 178 | 162 | 175 |
| | | 16.7% | 28.8% | 26.2% | 28.3% |

Table 5: Central-tendency and variability measures for the eight JCR categories

| Category | Measures | $R_1$=2-JIF | $R_2$ | $R_3$ | $R_4$ | 2M-JIF | 5-JIF |
|---|---|---|---|---|---|---|---|
| Astronomy & Astrophysics | Median | 1.683 | 1.874 | 1.679 | 1.600 | 1.982 | 1.757 |
| | Mean | 3.070 | 3.407 | 3.551 | 2.868 | 3.947 | 3.180 |
| | Sd | 4.292 | 5.563 | 5.597 | 4.931 | 5.927 | 4.803 |
| Biology | Median | 1.540 | 1.505 | 1.553 | 1.624 | 1.851 | 1.719 |
| | Mean | 2.097 | 2.341 | 2.346 | 2.500 | 2.663 | 2.374 |
| | Sd | 2.115 | 2.293 | 2.488 | 2.897 | 2.843 | 2.390 |
| Ecology | Median | 1.829 | 2.343 | 2.421 | 2.425 | 2.586 | 2.250 |
| | Mean | 2.643 | 3.168 | 3.292 | 3.530 | 3.651 | 3.122 |
| | Sd | 2.681 | 3.056 | 2.858 | 3.444 | 3.480 | 2.871 |
| Engineering, Aerospace | Median | 0.549 | 0.623 | 0.737 | 0.672 | 0.764 | 0.654 |
| | Mean | 0.680 | 0.799 | 0.869 | 0.885 | 0.975 | 0.833 |
| | Sd | 0.605 | 0.762 | 0.787 | 0.880 | 0.848 | 0.727 |
| History & Philosophy of Science | Median | 0.442 | 0.446 | 0.500 | 0.588 | 0.705 | 0.553 |
| | Mean | 0.580 | 0.659 | 0.682 | 0.735 | 0.855 | 0.725 |
| | Sd | 0.603 | 0.694 | 0.642 | 0.672 | 0.702 | 0.632 |
| Mathematics, Interdisciplinary Applications | Median | 0.893 | 1.079 | 1.230 | 1.132 | 1.376 | 1.131 |
| | Mean | 1.108 | 1.291 | 1.435 | 1.593 | 1.730 | 1.394 |
| | Sd | 0.771 | 0.884 | 1.087 | 1.662 | 1.545 | 1.033 |
| Medicine, Research & Experimental | Median | 2.297 | 2.376 | 2.320 | 2.274 | 2.675 | 2.418 |
| | Mean | 3.033 | 3.476 | 3.121 | 3.291 | 3.804 | 3.337 |
| | Sd | 3.290 | 3.979 | 3.943 | 4.197 | 4.313 | 3.635 |
| Multidisciplinary Sciences | Median | 0.510 | 0.571 | 0.828 | 0.650 | 0.864 | 0.789 |
| | Mean | 2.313 | 2.461 | 2.471 | 2.521 | 2.705 | 2.866 |
| | Sd | 6.419 | 7.003 | 6.918 | 6.823 | 6.942 | 7.231 |

Sd: Standard deviation

Table 6: Central-tendency and variability measures for the aggregate data

| Measures | $R_1$=2-JIF | $R_2$ | $R_3$ | $R_4$ | 2M-JIF | 5-JIF |
|---|---|---|---|---|---|---|
| Median | 1.245 | 1.442 | 1.431 | 1.478 | 1.745 | 1.531 |
| Mean | 2.142 | 2.453 | 2.449 | 2.538 | 2.827 | 2.481 |
| Within-group variance ($Sd^2$) | 3.203 | 3.717 | 3.670 | 3.800 | 3.998 | 3.505 |
| Between-group variance ($Sd^2$) | 0.709 | 0.790 | 0.717 | 0.729 | 0.795 | 0.728 |
| Reduction in the variance | 2.494 | 2.927 | 2.953 | 3.071 | 3.203 | 2.777 |

Sd: Standard deviation; Within-group: within the 618 journals; Between-group: between the JCR categories